\documentclass[lettersize,journal]{IEEEtran}
\IEEEoverridecommandlockouts
\usepackage{cite}
\usepackage{amsmath,amssymb,amsfonts}
\usepackage{algorithmic}
\usepackage{graphicx}
\usepackage{textcomp}
\usepackage{xcolor}

\def\BibTeX{{\rm B\kern-.05em{\sc i\kern-.025em b}\kern-.08em
    T\kern-.1667em\lower.7ex\hbox{E}\kern-.125emX}}
   
   \usepackage[version=3]{mhchem} 
\usepackage{chemarr}

     \makeatletter
\newcommand\xleftrightarrow[2][]{%
  \ext@arrow 9999{\longleftrightarrowfill@}{#1}{#2}}
\newcommand\longleftrightarrowfill@{%
  \arrowfill@\leftarrow\relbar\rightarrow}
\makeatother

\usepackage{widetext}

\begin{document}

\title{Further Experimental Evidence of the \emph{Dead Matter Has Memory} Conjecture in Capacitive Devices}

\author{
\IEEEauthorblockN{Anis Allagui$^{1,2}$, Di Zhang$^1$ and Ahmed Elwakil$^{3,4}$, \emph{Senior Member IEEE}}

\IEEEauthorblockA{\small
{$^1$Dept. of Sustainable and Renewable Energy Engineering, University of Sharjah, United Arab Emirates (aallagui@sharjah.ac.ae)} \\ 
{$^2$Dept. of Mechanical and Materials Engineering, Florida International University, Miami, FL33174, USA}\\
{$^3$Dept. of Electrical Engineering and Computer Science, Khalifa University,  United Arab Emirates (elwakil@ieee.org)}\\
{$^4$Dept. of Electrical and Software Engineering, University of Calgary, Alberta, Canada (ahmed.elwakil@ucalgary.ca)}
}
}

\maketitle

\begin{abstract}
This study provides new sets of experimental results supporting  Westerlund's conjecture that \emph{Dead Matter Has Memory} \cite{westerlund1991dead}.  Memory effects in the dynamic response of electric double-layer capacitors (EDLCs) that integrate its prior history of stimulation and state have been experimentally observed and reported in a few recent studies. The different excitation signals used to quantify such effects in these studies aimed at  charging a device to the same voltage value and the exact same accumulated charge level but in different manners. Having reached the same unique voltage-charge point,  it was observed that different yet repeatable discharge patterns occur, proving the existence of memory. The aim of this work is to provide further experimental evidence of the inherent memory effect in EDLCs in response to time-varying stationary input excitations with different statistical properties. In particular, different sets of charging voltage waveforms composed of fixed dc values with superimposed  uniformly-distributed random fluctuations of different amplitudes were created and used to charge the same EDLC device to a unique voltage-charge point. The duration of these signals was the same but with different values of variance around the mean value. We observed different time-charge responses depending on the extent of the noise level in these charging waveforms. This is   interpreted and discussed in the context of inherent memory using  fractional-order voltage-charge equations of non-ideal capacitors.

\end{abstract}
\



\section{Introduction\label{sec:Introduction}}

Over 30 years ago, Westerlund \cite{westerlund1991dead} highlighted the memory effect in non-ideal capacitors that follow  power-law (dis)charge dynamics    giving rise to the conclusion that dead matter must have memory. In a follow-up article, Westerlund and   Ekstam \cite{Westerlund-2} established the mathematical foundation of capacitor theory in the context of fractional-order calculus which cannot be avoided for the accurate modeling of non-ideal capacitive devices. The booming interest in fractional calculus and fractional-order dynamics has enabled more recent in-depth studies on the memory effect in dielectrics \cite{uchaikin2009memory}, neuron models \cite{teka2017fractional, teka2014neuronal} and supercapacitors \cite{uchaikin2016memory}. The experimental work done using supercapacitors in particular has provided solid proof of the existence of memory leaving no doubt to the validity of the \emph{dead matter has memory conjecture} \cite{memoryAPL}.
Supercapacitors are electrochemical  energy storage devices known for their outstanding power performance, excellent reversibility and long cycle life, but they are still lagging behind in terms of energy density when compared with batteries. Their charge storage mechanisms can be either physical, i.e. purely electrostatic ion adsorption in the electric double-layer at the electrode/electrolyte interface, or pseudocapacitive via highly reversible and relatively fast redox reactions, or both. Furthermore, the transport dynamics in such devices in response to an external excitation is mainly subdiffusive \cite{metzler2000random}, which can be attributed to several factors including  the {porous nature} of the electrodes, their fractal geometries and also  because of {internal friction forces and continuous scattering} of ions  diffusing in the supporting electrolytes and through the wetted pores \cite{ribeiro2016active}.   In addition,   different types of {inter-ion interactions, crowdedness and buffering in narrow pores}  play a  role in establishing  retarded diffusion processes in these  devices \cite{metzler2000random}. From a system-level perspective, the dynamics of supercapacitors follow a power-law behavior which can be viewed as the collective result of many coupled processes with different widely distributed time constants  \cite{10.1149/1945-7111/ac621e}. Multiple time scale dynamics hint at the existence of  memory effects that take into account the global, non-local time behavior of past activities up to the present time. 
This type of behavior  can be well described with fractional-order integro-differential equations \cite{du2013measuring, N1}.

In recent years, there has been an intense effort to experimentally verify and quantify the memory effect in EDLCs. For example, in \cite{memoryAPL} it was shown that discharging a supercapacitor  into a constant resistor from the same \emph{voltage and charge} point reached using two different charging waveforms (step voltage and linear voltage ramp) of two different durations leads to two different responses, mostly in the short term, transient regime. This indicated that contrary to ideal capacitors, knowledge of the initial condition  is not sufficient to determine the present and future states. 
 In a second study \cite{memQ},  a quantitative estimate of memory  using the voltage memory trace interpretation of fractional-order dynamics  \cite{teka2014neuronal} was provided. A follow-up paper by the same group illustrated the application of the memory effect  by sequentially encoding information into the charging pattern of the device, and then uniquely retrieving the coded information from the discharge response \cite{allagui2021possibility}.

In this work, we further investigate the memory effect  of  supercapacitors, albeit in response to time-varying stationary voltage excitations. The input voltage waveforms are  
 derived from uniform distributions of different sample ranges, means and variances.  
 It is observed that the extent of the noisy excitation around the mean value (i.e.  the prehistory and features of the charging sequence) determine the actual response of the device at any given time. 
 The cross correlation function between charge and voltage shows a visible degree of correlation, and thus short term memory effects,  that fade out as the time lag is increased. We also  allude to the effect of noise variance on the weighted sum memory trace that takes into account all prior history of the device. It is worth mentioning that in a recent study \cite{aip} aimed at explaining the \emph{cardiac memory} where the electrical activity in the heart depends on the prior history of one or more system state variables the following was mentioned:  \emph{$\ldots$ Memory is represented via capacitive memory, due to fractional-order  that arises due to non-ideal behavior of membrane capacitance}. Therefore,  we believe that studying the memory effect in capacitive devices is of significant importance.

\section{Theory}


Consider a linear capacitive device characterized in the frequency domain (steady-state condition) by a capacitive system function $C(s)$ ($s=j 2\pi f$ with $f$ being the frequency), such that  the charge-voltage relationship is given by:
\begin{equation}
 Q(s)=C(s)  \, V(s)
 \label{eq3}
 \end{equation} 
 The corresponding time-domain charge-voltage relationship would then be represented in terms of a hereditary (convolution) integrals  as \cite{nigmatullin1984theory, baleanu2010newtonian}:
\begin{align}
q (t) &=  \int_0^t c(t-\tau) v(\tau) d \tau = (c \circledast v)(t) \nonumber \\
&=\int_0^t v(t-\tau) c(\tau)  d \tau = (v \circledast c)(t)
\label{eq5}
\end{align}
where the lower limit of the integrals  can also be taken to be $-\infty$ to reflect infinite past. 
Here in   Eqs.\;\ref{eq3} and \;\ref{eq5}, $f(t)\xleftrightarrow[]{\text{\,}} F(s)$ represents the pair of  time-domain function $f(t)$ and its frequency-domain Laplace transform $F(s)$, assumed to exist, i.e. $\mathcal{L}\{f(t)\}(s) = F(s) = \int_{-\infty}^{\infty} f(t) e^{- s t} dt$ and 
$\mathcal{L}^{-1}\{F(s)\}(t) = f(t) = (2\pi j)^{-1} \int_{a-j\infty}^{a+j\infty} F(s) e^{s t} ds$. Analysis and discussions on these formulas and others for describing non-ideal capacitive devices can be found for instance in some of our recent work \cite{allagui2022time, ieeeted,allagui2021inverse}.  
The integral in Eq.\;\ref{eq5} indicates that 
the accumulated charge in the device at an instant $t$ depends on the values of the applied voltage taken at times in the interval $\tau \in [0,t]$ weighted by the convolution kernel  $c(t-\tau)$ characteristic of the device under study.

For the particular case of an ideal capacitor  we write 
$c(t-\tau) = C_1$ 
 where $C_1$ is a constant, independent of time or frequency. 
 Thus, Eq.\;\ref{eq5} simplifies to be the well-known  relation 
 \begin{equation}
\Delta q (t) = C_1 \Delta v (t)  
 \label{eq7}
 \end{equation} 
where $\Delta q (t)$ is evidently equal to $  q (t) - q(0) $, where $q(0)$ is an initial arbitrary charge stored on the device. The same is for the voltage difference $\Delta v (t)$. Ideal capacitors are therefore said to be memoryless devices.


However, if the convolution kernel $c(t-\tau)$ is not a constant, then all events over the history of the device contribute to its current state of charge.    
The capacitance of electrochemical capacitors  including EDLCs, pseudocapacitors and hybridized versions of the two cannot be taken as a constant. This is due to their complex structures, inhomogeneities and porosities of their electrodes, leading to distributed response times. These devices have been shown to exhibit fractional-order behavior \cite{uchaikin2009memory, uchaikin2016memory, N1, N2} that requires  for their frequency-domain   modeling invoking a fractional-order   impedance  function, e.g.:  
\begin{equation}
Z_{\text{CPE}} = \frac{V(s)}{I(s)}=\frac{1}{C_{\alpha}s^{\alpha}}
\label{eqZcpe}
\end{equation} 
 Note that $s^{\alpha}=\omega^{\alpha}\angle\,\alpha\pi/2$, with $0<\alpha \leqslant 1$ and $C_{\alpha}$ is in units of F\,s$^{\alpha-1}$ \cite{8301600}. The fractional-order capacitor is also known as the constant phase element (CPE) and its impedance reduces to that   of an ideal capacitor $1/(C_1 s)$ when $\alpha=1$. 
 The use of a fractional-order model indicates \textit{a priori} that memory effects should be expected in time-domain experiments. We note in this regards that  the time-domain voltage-current relationship  corresponding to Eq.\;\ref{eqZcpe} is given,  by inverse Laplace transform, by the integro-differential equation:
\begin{equation}
i (t)=C_{\alpha}\, _{0}^C\mathrm{D}_{t}^{\alpha}  v(t)
\end{equation}
instead of the traditional ODE, $i (t)=C_1 dv (t)/dt$, known for ideal capacitors.  The operator $_{0}^C\mathrm{D}_{t}^{\alpha}$  represents here the right-sided Caputo time fractional derivative of order $\alpha >0$,  
 \begin{equation}
  _{0}^C\mathrm{D}_{t}^{\alpha}f(t) = \frac{1}{\Gamma{(m-\alpha)}} \int_{0}^{t} {(t-\tau)^{-\alpha-1+m}} {f^{(m)}(\tau)}  d\tau
  \label{eq1}
\end{equation} 
where $m-1 < \alpha \leqslant m$, $ m \in \mathbb{N} $, and ${f^{(m)}(\tau)}$ is the $m^{th}$ derivative of ${f(\tau)}$ (in our case $m=1$).  Again, this type of integrals indicate that all prior history of the function should be accounted for in order to determine the actual behavior of the device at the present time\;$t$. 
This can also be viewed from the finite difference approximation of Eq.\;\ref{eq1}. 
    For $t_k = k\Delta t$, $k=0,1, \ldots, K$ and using a time step $\Delta t = T/K$, then for all  $0 \leqslant k \leqslant K-1$ we have:
\begin{align}
_{0}^C\mathrm{D}_{t}^{\alpha}&  v(t_{k+1}) \approx \frac{1}{\Gamma(2-\alpha)}     
\sum\limits_{j=0}^k b_j 
\frac{\left[ 
v (t_{j+1})- 
v (t_{j} )
\right]}{\left(\Delta t\right)^{\alpha}}  
\label{eq:LX1}
\end{align}
where $b_j=\left[  
(k+1-j )^{1-\alpha} 
- (k-j)^{1-\alpha} 
\right]
$, with a $(2-\alpha)$-order accuracy in time   \cite{lin2007finite}. 
We can  rewrite Eq.\;\ref{eq:LX1}  as:
\begin{align}
_{0}^C\mathrm{D}_{t}^{\alpha} v(t_{k+1}) 
 \approx 
 \frac{ (\Delta t)^{-\alpha}}{  \Gamma (2-\alpha)} &   \left\{ 
v(t_{k+1}) - v(t_{k})  \right.  \nonumber +\\
&  
 \sum_{j=0}^{k-1} 
b_j \left.
\left[ v (t_{j+1})- v (t_{j} )\right]  \right\}
\label{eq:memtr1}
\end{align}
which is the the sum of the immediate past of the voltage (first two terms, weighted by the Gamma function and the fractional time, $(\Delta t)^{\alpha}$) and a memory trace \cite{teka2014neuronal}   (summation term) that contains voltage information about all previous activity of the device.  The   memory trace term increases when the fractional order  $\alpha$ in  Eq.\;\ref{eq:memtr1} decreases further away from one \cite{teka2017fractional,memQ}. It is understood  that at the limiting case of $\alpha=1$ (corresponding to first  order integer derivative), the behavior of the differentiable function is determined only locally in an infinitesimal neighborhood of the considered  point. In this case, the memory trace part vanishes, and does not have any effect on the dynamics of the device \cite{teka2017fractional}, which is simply a memoryless capacitor.  


Referring back to the convolution integrals and their system function kernels, the  expression of the current represented by 
Eq.\;\ref{eq1} with $m=1$ (i.e. $0< \alpha \leqslant 1 $) 
can be written as the Laplace convolution:
\begin{equation}
i(t) = \frac{ C_{\alpha} t^{-\alpha}}{\Gamma{(1-\alpha)}}  \circledast \frac{dv(t)}{dt}
\end{equation}
where the capacitive kernel is the power law function
  \begin{equation}
c(t) = \frac{ C_{\alpha} t^{-\alpha} }{\Gamma{(1-\alpha)}}  
\end{equation}
that tends to zero as $t\to\infty$.  
The charge-voltage relationship for these   non-ideal capacitive devices in response to a voltage signal $v(t)$   is not simply $\Delta q (t) = C_1 \Delta v (t)  $ (Eq.\;\ref{eq7}), but rather: 
\begin{equation}
q(t) = \frac{ C_{\alpha} t^{-\alpha}}{\Gamma{(1-\alpha)}}  \circledast {v(t)}
\end{equation}
or in discrete form:
 \begin{align}
\Delta q (t)  = \frac{C_{\alpha} }{\Gamma(1-\alpha)} \sum\limits_{j=0}^k (t-t_{j})^{-\alpha} v(t_j)
\label{eq10}
\end{align}
This relation clearly indicates the presence of a memory effect in these devices \cite{teka2014neuronal}.

\section{Experiment}

\begin{figure}[!h]
	\begin{center}
		\includegraphics[width=.35\textwidth]{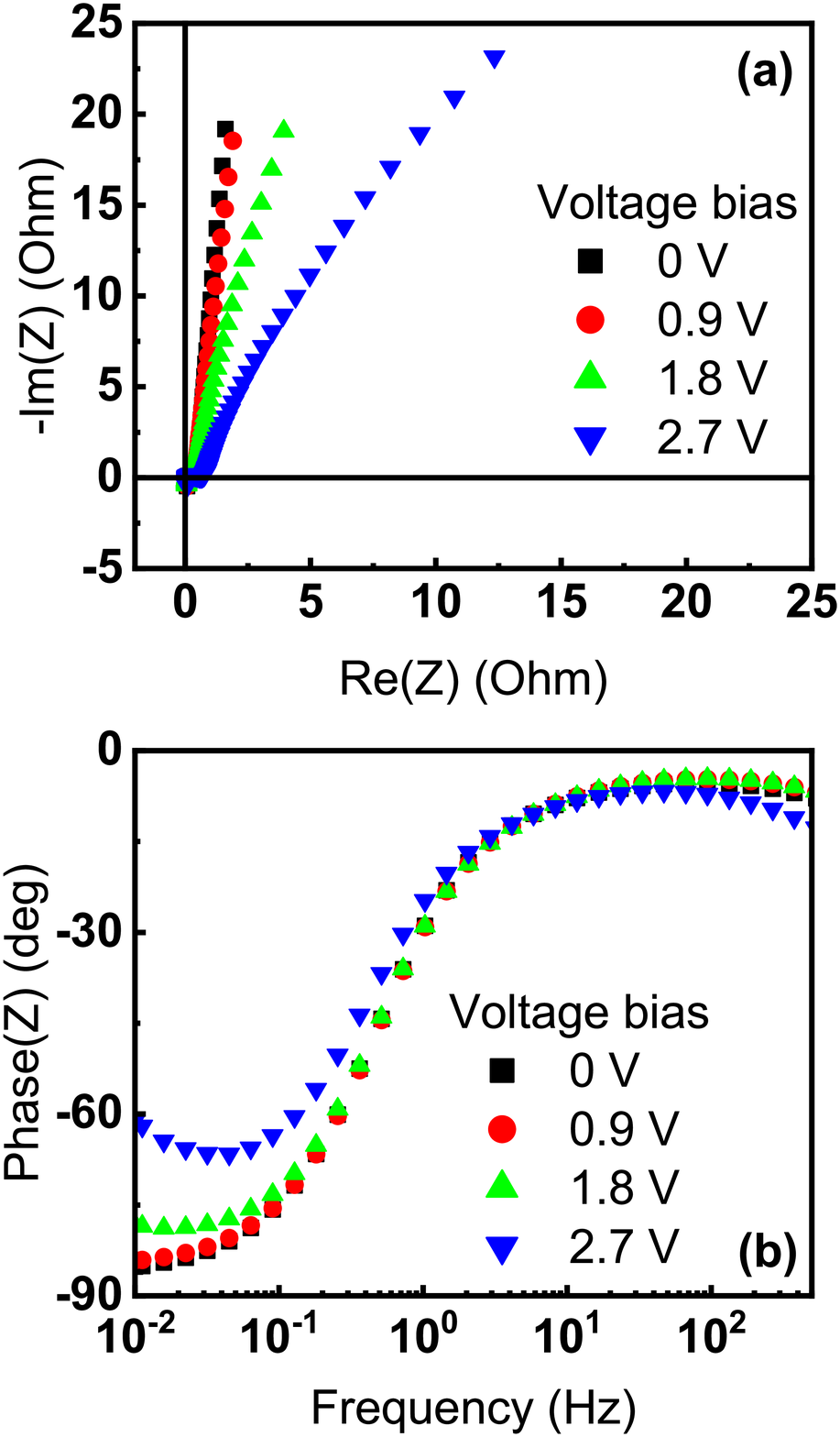}
						\caption{Impedance spectroscopy results of the GHC NanoForce supercapacitor (rated   1\,F, 2.7\,V). Measurements were carried out on a Biologic VSP-300 electrochemical station using  traditional stepped sine excitations of frequencies  10\,mHz  to 100\,kHz with an amplitude of 10\,mV around a given dc voltage, and  the data were recorded at 10 frequency points per decade.}
		\label{fig11}
	\end{center}
\end{figure}


The device under test for this study is a commercially-available GHC NanoForce supercapacitor, rated 1\,F, 2.7\,V, and the electrical measurements were  conducted  on a Biologic VSP-300 electrochemical station equipped with an impedance spectroscopy module\footnote{ This  instrument has a voltage resolution of 1$\mu$V on 60\,mV,  voltage accuracy  $< \pm$1 mV, and maximum voltage scan rate of 0.2\,V/ms. Its current range  is 10\,nA to 1\,A with lowest accuracy and  resolution of   $\pm$100\,pA and 0.8\,pA, respectively, on a 10\,nA scale. }. 

Before we analyze time-domain data, we first characterized the device by impedance spectroscopy. 
The results in terms of Nyquist  representation of real vs. imaginary parts of impedance, and impedance phase angle vs. frequency at 0, 0.9, 1.8 and 2.7\,Vdc are shown in Fig.\;\ref{fig11}. The figure shows a clear deviation  of the spectral impedance  of the device from that of  an ideal capacitor. This impedance can be fitted good enough with an $R_s$-CPE model \cite{N1,memoryAPL, allagui2021possibility}, which again suggests memory effects  in time-domain measurements.

\begin{figure*}[!h]
	\begin{center}
		\includegraphics[width=.95\textwidth]{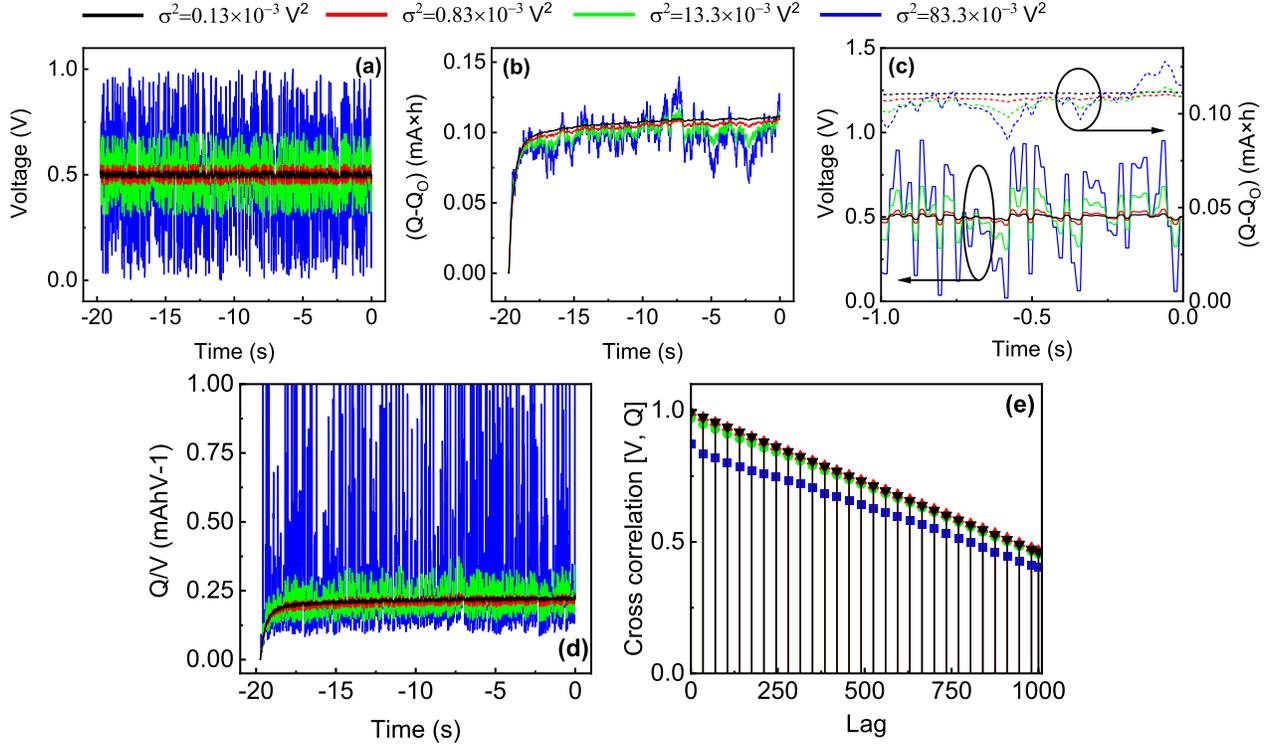}
		\caption{(a) Applied time-domain voltage signals with a mean value of 0.5\,Vdc and (b) accumulated charge on the GHC NanoForce supercapacitor (rated   1\,F, 2.7\,V) when excited with these waveforms. (c) The synchronized voltage and charge vs. time for the last second of the experiment. (d) Plot of the ratio $\Delta q /\Delta v$ vs. time and (e) plot of $R_{vq}(\tau)$ (Eq.\;\ref{eqR}) vs. number of time lags}
		\label{fig1}
	\end{center}
\end{figure*}

Secondly,  the supercapacitor was charged using  excitation voltage signals of the type 
\begin{align}
v (t) = v({t_0}) H(t) \nonumber + [v({t_1})-v({t_0})] H(t-t_1)  \nonumber 
 +\ldots \nonumber \\
 + [v({t_j}) - v({t_{j-1}})] H(t-t_j)
\label{eqvc}
\end{align}
 where $H(t-t_k)$ is the Heaviside step function. We used  64 different settings for $v(t)$ by full factorial of (i) four  different voltage  mean values ($V_{dc}=\mu=\text{0.5}$, 1.0, 1.5 and 2.0\,V) with (ii) four different ranges of uniformly-distributed voltage fluctuations  (i.e.   $\pm$20, $\pm$50, $\pm$200 and $\pm$500\,mV) superposed on the dc values, and (iii) four different durations ($\Delta t=\text{1}$, 5, 20 and 60\,seconds). 
It is clear that a uniform noise 
means the signal consists of values derived from a uniform distribution of probability density function expressed as
 \begin{equation}
f(x)  
=\left\{ 
\begin{aligned} 
   &\frac{1}{ b-a}, \;\; a   \leqslant x \leqslant b &  \\
  & 0, \;\; \text{elsewhere}&
\end{aligned} 
\right.
\end{equation}
 with variance $\sigma^2=(b-a)^2/12$ taking the values of (0.13, 0.83,   13.3, and   83.3)$\times$10$^{-3}$\,V$^2$ for the  ranges  $\pm$20, $\pm$50, $\pm$200 and $\pm$500\,mV, respectively.  
Each signal consisted of $N=\text{1000}$ points equispaced in time with a time step of $\Delta t=  T/N$.   
 Prior to each applied excitation, the supercapacitor was  discharged down to 5\;mV using a 10\;$\Omega$ resistor.

\section{Results and Discussion}

Typical results depicting the memory effect observed with the GHC NanoForce  supercapacitor  are shown in  Fig.\;\ref{fig1}, corresponding to  20-seconds long, 0.5\;V  mean value voltage  excitations with   four different superposed uniformly-distributed fluctuations (see Fig.\;\ref{fig1}(a)). The last value of each charging sequence is appended to 0.5\;V. The results obtained with the other dc voltage values and   other time durations  are similar to those in Fig.\;\ref{fig1}, and  not shown here to preserve space. \
The accumulated charges in mA.h in response to the voltage excitations of Fig.\;\ref{fig1}(a) are shown in  Fig.\;\ref{fig1}(b). 
The overall trends of the  curves start first with a relatively quick power-law  growth of the accumulated charge followed by an asymptotic limit as the charging time is increased \cite{allagui2021possibility}. 
We also observe from this quasi-stationary signal that in response to a sudden change of a voltage stimulus, the device responds accordingly with relatively large change in the instantaneous charge, followed by slower adaptation  until the next change in voltage input  (see Fig.\;\ref{fig1}(c)). The overall extent of change in the accumulated charge superposed on the general trend is accentuated by the increase in variance of the input voltage signal (Fig.\;\ref{fig1}(b) and Fig.\;\ref{fig1}(c)). This leads to the primary important conclusion  that the device's accumulated charge adapts to changes in the stimulus statistics, not only in terms of its mean value but also in its variance. 

Plots of the point-by-point ratios of charge to voltage are shown in Fig.\;\ref{fig1}(d). These results show also that by the end of the charging period, the overall accumulated charges are  different from each other even though the last value of voltage was set to 0.5\;V in all cases\footnote{The same is observed for the other tested scenarios, i.e. with different mean values and different excitation durations.}.  

In Fig.\;\ref{fig1}(e), plots of normalized cross-correlations between voltage and charge   for the different levels of noise used  are shown. The  cross-correlation is defined as 
$R_{vq}(\tau) = E \left\{  v (t) \, q (t+\tau) \right\}$
where $E \left\{  \cdot \right\}$ denotes the expected value. The charge-voltage correlation as a function of time lag   when the dc voltage input signal is superposed with the highly fluctuating signal of $\sigma^2=\text{83.3} \times\text{10}^{-3}$\,V$^2$  always shows lower values compared to the cases with lower variance values.  This indicates a lack of correlation between   the two variables and thus the invalidity of the constitutive relation  $\Delta q (t) = C_1 \Delta v(t)$, which applies for the case of ideal capacitors only \cite{ieeeted}. Furthermore, the correlation coefficient defined as 
$ \rho_{vq} = {  \mathrm{Cov}\{ v  , q \}  }/{ { \sigma_{v }  \sigma_{q }}} $ 
where 
 $\mathrm{Cov} \left\{  \cdot \right\}$ denotes the covariance between $v  $ and $ q  $ are found to be -0.0491,  -0.0297, 0.0684 and 0.2141 for the four different  superposed signals  respectively. The highest value of $\rho_{vq}$ goes with the highest variance of noise. This is an indicator that the device remembers the extent of the noise level.

\begin{figure}[!h]
	\begin{center}
		\includegraphics[width=.35\textwidth]{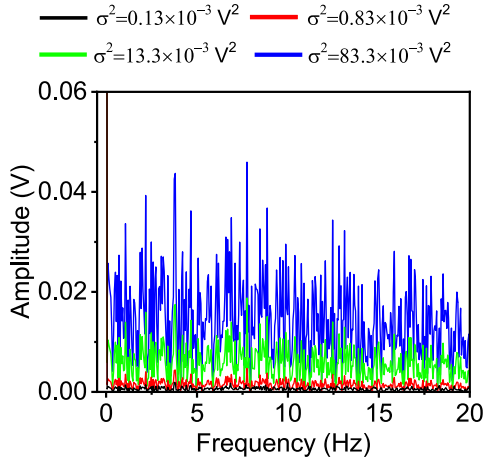}
		\caption{Discrete-time Fourier transform (DTFT) of the applied charging voltage excitations depicted in Fig.\;\ref{fig1}(a)}
		\label{fig3}
	\end{center}
\end{figure}

\begin{figure}[h]
	\begin{center}
		\includegraphics[width=.4\textwidth]{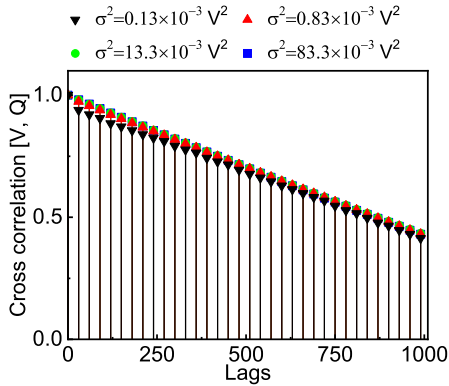}
		\caption{Normalized cross-correlations between voltage and charge vs. number of time lags. The charge is  obtained using Eq.\;\ref{eq10} with $C_{\alpha}=\text{0.660 F\,sec}^{\alpha-1}$, $\alpha=\text{0.905}$, and  with the same voltage signals applied experimentally on the supercapacitor device as depicted in Fig.\;\ref{fig1}(a)}
		\label{Di}
	\end{center}
\end{figure}

In connection with Eq.\;\ref{eq10} which indicates that the memory trace term is more pronounced when the value of $\alpha$ decreases, we note that  the magnitude of non-dc frequency components of the  voltage input $v(t)$   increase with the increase of variance of the uniform noise (see DTFT  of voltage excitations  in Fig.\;\ref{fig3}). The stronger high-frequency components in the excitation signal are directly related to the increase of impedance phase angle in the device (i.e. lower  value for $\alpha$), as depicted in Fig.\;\ref{fig11}(b). Hence, the higher are the magnitudes of these frequency components, the stronger is the effect of the memory trace term in Eq.\;\ref{eq10}. To further verify this point,  Eq.\;\ref{eq10} was numerically executed with the same voltage signals applied experimentally on the supercapacitor device (Fig.\;\ref{fig1}(a)), and with $C_{\alpha}$ and $\alpha$ being 0.660 F\,sec$^{\alpha-1}$  and 0.905, respectively. These values are estimates of the CPE model parameters used to fit the impedance data measured at different dc voltage biases. 
 The normalized cross-correlations between these voltage signals   and their corresponding   charges obtained via Eq.\;\ref{eq10}   are plotted in Fig.\;\ref{Di}, and show   similar trends to the experimental results (see Fig.\;\ref{fig1}(e)). 

\section{Conclusion}

In summary, we have provided further experimental evidence of the inherent memory effect in EDLCs which results from the complex multi-time scale dynamics of their internal microscopic structure and mechanisms. These results, in conjunction with those previously reported in \cite{allagui2021possibility}, show that it is possible to encode information in the statistical distribution of the random noise signal superimposed on the dc step charging voltage of a device. From the fractional calculus modeling  point of view, the memory effect  in these devices is attributed to stronger higher frequency harmonics of the input signal, and as a consequence to the extent of deviation of the device from ideality. In other words,   high variance fluctuations make the device operate in a less capacitive, more resistive mode with lower values of the fractional coefficient $\alpha$, and thus with a stronger memory trace that integrates all past values of its state.  We believe that there is no further doubt regarding the validity of the  "dead matter has memory" conjecture \cite{westerlund1991dead}. 

\bibliographystyle{IEEEtran}



\end{document}